\def\preprint{0}                
\def\preprint{1}                
\def\comment#1{}
\if\preprint1
        \documentclass[usenatbib]{mn2e}
        \usepackage{times}
        \usepackage{graphicx}
        \usepackage{calc}

\else
        \documentstyle[astrop-bib,referee,times]{mn}
        \newcommand{\includegraphics}[1]{}
\fi

\def\oversim#1#2{\lower0.5pt\vbox{\baselineskip0pt \lineskip-0.5pt
     \ialign{$\mathsurround0pt #1\hfil##\hfil$\crcr#2\crcr\sim\crcr}}}

\title[Mass-losing AGB stars in Sgr and Fornax]{Dust mass-loss rates from 
AGB stars  in the Fornax and Sagittarius dwarf Spheroidal galaxies }
\author[Eric Lagadec et al.]{Eric Lagadec$^{1}$\thanks{E-mail:
eric.lagadec@manchester.ac.uk}\thanks{ Based on observations collected at the
European Southern Observatory, Chile (ESO Programme 075.D-0443(A) )} 
Albert A. Zijlstra$^{1,2}$, 
Mikako Matsuura$^{3,4}$, 
J.W. Menzies$^2$, \newauthor Jacco Th. van Loon$^{5}$, Patricia A. Whitelock$^{2,6}$ \\
$^{1}$ Department of Astronomy, University of Manchester, Sackville street, Manchester M601QD, UK \\
$^{2}$ South African Astronomical Observatory, PO Box 9, 7935
Observatory,  South Africa\\
$^3$  Division of Optical and IR Astronomy, National Astronomical Observatory 
     of Japan, Osawa 2-21-1, Mitaka, Tokyo 181-8588, Japan\\
$^4$Department of Physics and Astronomy
University College London
Gower Street, London WC1E 6BT, UK \\
$^5$ Astrophysics Group, School of Physical \&\ Geographical Sciences, 
Keele University, Staffordshire ST5 5BG, UK\\
$^6$ National Astrophysics and Space Science Programme, Department of
Mathematics and Applied Mathematics, and Department of Astronomy, \\University
of Cape Town, Rondebosch, 7701, South Africa}

\begin{document}

\date{Accepted . Received 1988}

\pagerange{\pageref{firstpage}--\pageref{lastpage}} \pubyear{2002}

\maketitle

\label{firstpage}

\begin{abstract}
 To study the effect of metallicity on the mass-loss rate of asymptotic
giant branch (AGB) stars, we have conducted mid-infrared photometric
measurements of such stars in the Sagittarius (Sgr dSph) and Fornax dwarf
spheroidal galaxies with the 10-$\mu$m camera VISIR at the VLT.  We derive
mass-loss rates for 29 AGB stars in Sgr dSph and 2 in Fornax.  The dust
mass-loss rates are estimated from the $K-[9]$ and $K-[11]$ colours. 
Radiative transfer models are used to check the consistency of the method. 
Published IRAS and Spitzer data confirm that the same tight correlation
between $K-[12]$ colour and dust mass-loss rates is observed for AGB stars
from galaxies with different metallicities, i.e. the Galaxy, the LMC and the
SMC.

 The derived dust mass-loss rates are in the range
5$\times10^{-10}$ to 3$\times10^{-8}$ M$_{\odot}$yr$^{-1}$ for the observed AGB
stars in Sgr dSph and around 5$\times10^{-9}$ M$_{\odot}$yr$^{-1}$ for those in
Fornax; while values obtained with the two different methods are of the same
order of magnitude.  The mass-loss rates for these stars are higher than the
nuclear burning rates, so they will terminate their AGB phase by the
depletion of their stellar mantles before their core can grow significantly. 
Some observed stars have lower mass-loss rates than the minimum value predicted by theoretical models.

\end{abstract}

\begin{keywords}
circumstellar matter -- infrared: stars.
\end{keywords}

\section{Introduction}

Stars with initial mass in the range 0.8--8 M$_{\odot}$ go through an
asymptotic giant branch (AGB) phase towards the end of their evolution.  This
evolutionary phase is dominated by strong mass loss.  The star expels
material at rates up to $10^{-4}M_{\odot}$yr$^{-1}$, eventually ejecting between
20 and 80 per cent of its initial main-sequence mass. This process is of great
importance for the chemical evolution of our Galaxy. Mass loss from AGB stars
contributes to around half of the gas recycled by stars (Maeder 1992), and is
a dominant source of Galactic dust.

The mass-loss mechanism of AGB stars is a two-step process. First, shocks
due to pulsations from the star produce gas over-densities in an extended
atmosphere (e.g. H\"ofner et al. 1998). This triggers the formation of dust. Secondly, radiation
pressure accelerates the dust to the escape velocity. The gas is carried
along through friction with the dust particles. Pulsations alone can
explain mass-loss rates up to about $10^{-7}M_{\odot}$yr$^{-1}$, but the much
higher rates observed require dust-driven winds (Bowen \&\ Wilson 1991).

The importance of metallicity on the mass-loss rates of AGB stars is not
well understood. In low metallicity environments less dust is expected to
form, leading to lower predicted mass-loss rates.  Theoretical work by Bowen
\& Willson (1991) predicts that for metallicities below [Fe/H]$=-1$
dust-driven winds fail, and the wind becomes pulsation-driven. This would
affect the evolution of a low metallicity host galaxy in two obvious ways.
First the stellar dust production would be much reduced and therefore the
composition of the material out of which new stars and planets were forming
would be significantly different. Secondly the much weaker dust-driven
winds allow the degenerate core of the AGB star to grow for longer,
resulting in much higher masses for the remnant white dwarfs. In extreme
cases, the core could reach the Chandrasekhar limit before mass loss
terminates its evolution. The Bowen \&\ Willson mass-loss rates predict the
occurrence of AGB supernovae at very low metallicities (Zijlstra 2004). It is
therefore important to test whether dust-driven winds exist at low
metallicity.

Observations in the Magellanic Clouds and the Galaxy have shown that the
dust mass-loss rates are smaller in the Magellanic Clouds.  Assuming that
the dust-to-gas ratio is a linear function of metallicity ([Fe/H]=$-0.6$
(Venn 1999) and [Fe/H]=$-0.3$ for the Small and Large Magellanic Clouds
respectively) yields the conclusion that the total mass-loss rate
(dust+gas) may be similar in the three galaxies (van Loon 2000, 2006), although this
assumption may not be strictly correct, as the dust in the carbon stars
comes from primary carbon.  In order to obtain further constraints on the
effect of metallicity on the mass-loss rates from AGB stars, and to know if
dust-driven mass loss can occur at very low metallicities, we need to study
the mass-loss rates from more metal-poor AGB stars.

The dusty circumstellar envelopes surrounding these AGB stars absorb the
radiation from the central star and re-emit it in the thermal
infrared. Observations at infrared wavelengths have enabled the study of
mass-loss from Galactic and Magellanic Clouds AGB stars. Thanks to the
sensitivity achievable with mid-infrared instruments on 8m class ground-based
telescopes, it is now possible to perform such studies for AGB stars in more
distant Local Group galaxies.

We thus observed AGB stars in the Fornax dSph (Fornax) and Sagittarius (Sgr dSph) dwarf spheroidal
galaxies using the mid-infrared camera VISIR on the VLT (ESO, Chile). 
Fornax is a satellite galaxy of the Milky Way, at a distance of $\sim$ 138
kpc. Sgr dSph is located behind the Galactic disc and bulge at a distance of
$\sim$24 kpc.  It is currently being torn apart by tidal forces (Majewski et al. 2003). Those
galaxies have low metallicities (see section 2). Furthermore the distances
of both galaxies are quite well known, making an estimation of the mass-loss
rates easier than it is for stars within the Galaxy.

\section{Observations and target selection}

\begin{figure}
\includegraphics[width=\columnwidth,clip=true]{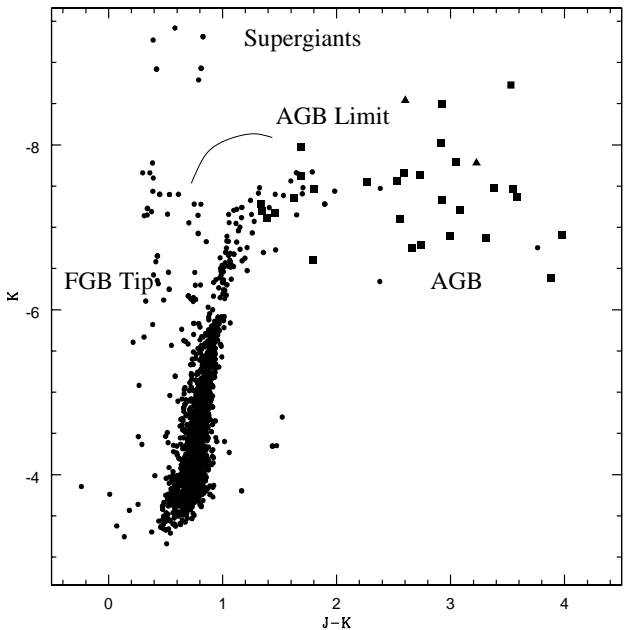}
\caption{\label{JHK} The $M_{K_s}$,$J-K$ diagram of the observed sample.
Squares represent the Sgr dSph stars and triangles the Fornax stars. Circles
show the stellar population of Fornax (See Section \ref{tar_sel} for more details).}  
\end{figure}
\subsection{Target galaxies}

We selected targets in two satellite galaxies of the Milky Way: the
Sgr dSph and the Fornax Dwarf Spheroidal
galaxy. Their distance moduli are taken as 17.02 mag for Sgr dSph (Mateo et al. 1995)
and 20.66 mag for Fornax (Bersier 2000).

Sgr dSph is located behind the Galactic Bulge. It is a substantial galaxy, but
it is being disrupted by the Milky Way: the tidal tails surround the Galaxy. 
The majority of its stellar population has a metallicity in the range
[Fe/H]\,=\,$-0.4$ to $-0.7$ and an age of $8.0\pm1.5$\,Gyr (Bellazzini et al.
2006). Two of the four planetary nebulae in this galaxy have identical
abundances of [Fe/H]$=-0.55$ (Dudziak et al. 2000).  The carbon stars
discovered by Whitelock et al. (1996, 1999) are plausibly related to the same
population as the planetary nebulae.  A more metal-rich population with
[Fe/H]\,=$-0.25$ also exists (Zijlstra et al. 2006a; Bonifacio et al. 2004),
although it  may not be uniformly distributed over the galaxy. Finally,
there is a minor, very metal-poor, population with Fe/H]$\sim-2$ which is
seen in the globular clusters, but also in one tidal-tail planetary nebula.

Fornax shows similar mass and properties to Sgr dSph, but is too far from the
Galaxy to suffer obvious tidal disruption.  It also shows evidence for an
extended period of star formation with the metallicity increasing over time. The
younger population (at $\sim 3\,10^8$\,yr even younger than Sgr dSph) has a
metallicity of [Fe/H]\,$=-0.7$ (Saviane et al. 2000) to
$\sim -0.6$ (Pont et al. 2004). The dominant population has an age of
2--10\,Gyr and metallicity [Fe/H]\,$=-1.0$ (Saviane et al.)  or $-0.9$ (Pont
et al.). There is also a very metal-poor population as in Sgr dSph, with
[Fe/H]$\sim-2$. The most metal-rich population is found mainly in the central
regions. 

\subsection{Target selection}
\label{tar_sel}
The targets within Sgr dSph (this galaxy subtends an angle of $\sim$ 10$^o$ in the sky
 and its center coordinates are $\alpha(2000) =$ 18$^h$55$^m$04$^s$ and $\delta(2000)=$-30$^o$28$'$42$''$)) 
 were selected in
two ways. First, we used the list of carbon stars given by Whitelock et
al. (1999 --- their table 1) with preference being given to the Miras with
measured periods or to stars with near-infrared red colours; these are spectroscopically
confirmed carbon stars. These are the stars with WMIF names in column 3 of Table \ref{log_obs}.
 Secondly, stars were selected from the two micron all sky survey (2MASS) catalogue 
with the following properties: $8<K_s<12$, $2.6<J-K_s<4.5$,
$0.4(J-K_s)+0.25<J-H<0.56(J-K_s)+0.36$. This selection should isolate carbon stars
comparable to, or somewhat redder than, those discussed by Whitelock et al.
and are a subset of those that have been monitored by Menzies et al. (in
preparation) at $JHK_s$ with the 1.4m Infrared Survey Facility (IRSF) at
Sutherland. The carbon-rich nature of these stars has yet to be
spectroscopically confirmed. Indeed it is not possible to disentangle oxygen-rich from
 carbon rich red AGB stars from JHK colours alone but in a low metallicity environment
 like Sgr, we expect more AGB stars to be carbon rich than in the Galaxy. All the stars
 with WMIF names in Table \ref{log_obs} have been confirmed as carbon stars from
 spectroscopy (Whitelock et al. 1999) and we will assume that all the other observed
 stars are carbon-rich, keeping in mind that some might be oxygen-rich. We will assume [Fe/H]$\sim -0.55$, based on the
descendant PNe population (Zijlstra et al. 2006a) and the carbon-star
population (Whitelock et al. 1996, 1999), but a range of metallicities is probably
represented. de Laverny et al. (2006) find for two Sgr dSph carbon stars (not
observed by us) [M/H]$=-0.5$ and $-0.8$.

Two carbon stars in Fornax were observed: these have similarly red colours to
the Sgr dSph sample and are spectroscopically confirmed carbon stars (Matsuura et al. 2007).
 Both stars are located in the outer regions of Fornax
($7^\prime$ and 15$^\prime$ from the centre). We assume that they have
the [Fe/H]$\sim -0.9$ of the dominant population, as the more metal-rich
population is concentrated to the centre (Saviane et al. 2000).

We also observed the star IRAS 20176--1458, selected from Mauron et al. (2004).
This is a carbon star in the Galactic halo. All AGB carbon stars in the halo
are expected to have escaped from one of the satellite galaxies, and the
proximity of this star to the main body of Sgr dSph may indicate an association.
Assuming an absolute magnitude of $M_{\rm bol}=-4$, its distance is around
15\,kpc which is not inconsistent with the tidal tail of Sgr (Law et al.
2005). The star lies several degrees off the location of the tidal stream
(Majewski et al. 2003), but as noted by these authors, carbon stars may
trace a different tidal event than the M stars. Mauron et al. (2004) argue
against an association with Sgr dSph.

The interstellar reddening towards Fornax is quite small, $E(B-V)\sim0.03$, and thus
negligible in $K_s$ (Demers et al. 2002).  For Sgr dSph, the reddening is
estimated to be $E(B-V)\sim0.14$ (Layden \& Sarajedini 2000).  The reddening
law from Rieke \& Lebofsky (1985) yields (A$_J$,A$_H$,A$_K$) =
(0.13,0.09,0.05). The star Sgr01 is closer to the Bulge and may suffer a
higher extinction: the NED model indicates $E(B-V)=0.24$ for this position.

The observed stars in both galaxies are shown in the (de-reddened) $M_{K}$ \textit{vs}
$J-K$ diagram in Fig. \ref{JHK}. Squares represent stars in Sgr dSph and triangles stars
in Fornax.  To show that the programme stars are on the AGB, we over-plotted the
distribution of Fornax field stars: these data come from an unpublished survey
using the 1.4m IRSF at Sutherland, and refer to a large area
around the centre of Fornax.  The tip of the red giant branch (RGB) is clearly
visible, with a gap just above it showing the onset of hydrogen burning and
thermal pulses. 

Most of the selected stars have magnitudes consistent with thermal-pulsing AGB stars.
They show a similar range to the mass-losing AGB stars in the SMC (Lagadec et
al. 2007), but tend to be located towards the fainter end of those stars,
with only three stars above $M_{K} = -8$.

\subsection{VISIR observations}

The observations were made with the VLT spectrometer and imager for the
mid-infrared (VISIR, Lagage et al. 2004), located at the Cassegrain focus of
the Very Large Telescope (VLT) UT3 at Paranal, Chile. The settings of our
observations gave a $0.075''$ per pixel scale and a field of view of
$32.5''\times32.5''$. These observations were carried out in visitor mode
during 4 nights from the 24th to the 28th of July 2005.  We observed 29 AGB
stars in Sgr dSph, 2 in Fornax and 1 in the Galactic halo. Some observations
were repeated if the first measurement was made in poor seeing conditions.
The log of the observations is presented Table \ref{log_obs}.

The observations were affected by unstable weather: approximately half the
time was non-photometric and could not be used. Seeing as reported by the
optical seeing monitor varied between 0.4 and 3 arcsec.

To reduce the background emission from the sky and the telescope, we used the
standard mid-infrared technique of chopping and nodding. To avoid saturation
of the detector by the ambient photon background, each individual nod cycle
was split into many short exposures of $\sim$10ms each. This procedure was repeated
for as many cycles as needed to obtain sufficient signal-to-noise.

The data reduction was performed using our own IDL routines. Images
were corrected for bad pixels, and then co-added to produce a single
flat-field corrected image, comprising the average of the chop and nod
differences. Standard stars were observed and analysed in the same way to
flux-calibrate our observations.  The flux calibration was performed using 
standard aperture photometry methods, applied to the programme and
reference stars. 

We observed all the stars with the VISIR PAH1 filter ($\lambda_c$=8.59$\mu$m,
$\Delta \lambda=0.42\mu$m). The brighter stars were also observed with the
VISIR PAH2 filter ($\lambda_c$=11.25$\mu$m, $\Delta \lambda=0.59\mu$m).  The
filters were selected for their very good sensitivity: they avoid the telluric
ozone band.  The zero-points for the two filters are taken as 53.7 (8.59$\mu$m)
and 31.5 Jy (11.59$\mu$m) respectively (values taken for a blackbody with T$_{eff}$=10000K and using Vega as a reference).
 The magnitudes derived using these
zero-points are designated below as [9] and [11] respectively.

The measured magnitudes, together with the 2MASS or SAAO $JHK$ photometry, are
listed in Table~\ref{photo}. For comparison, we also list the IRAS [12]
magnitude, taken from the Faint Source Catalogue (FSC), available for six
sources.  The IRAS magnitude is not colour-corrected. The IRAS magnitudes are
in most cases a little brighter, by a few tenths of a magnitude.  The IRAS
filter extends to longer wavelengths than the VISIR filter, and the IRAS
detected stars are red. The photometry is accurate to better than $\pm$0.05 mag
 at JHK for the stars observed at SAAO and around $\pm$0.02-0.03 mag for the stars observed with 2MASS.

\begin{figure}
\includegraphics[width=\columnwidth,clip=true]{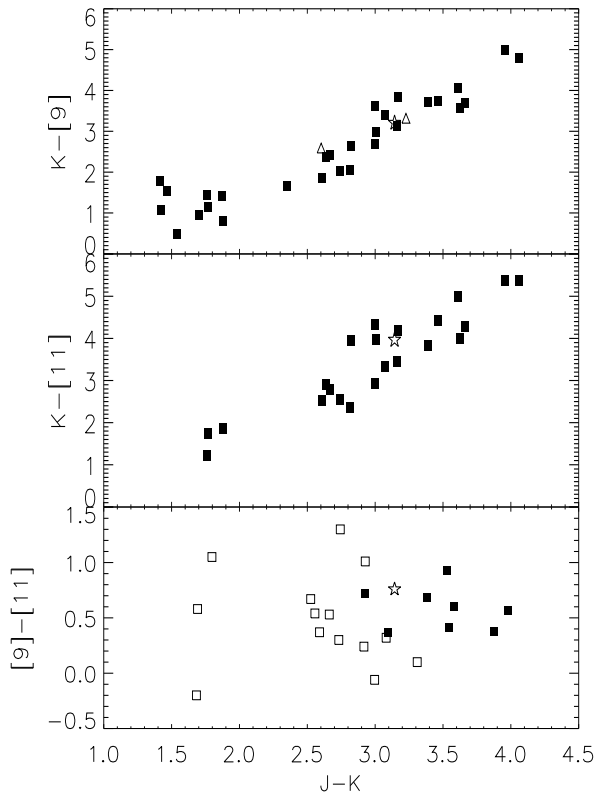}
\caption{\label{jk} The interstellar reddening-corrected
$J-K$ colours versus the mid-infrared excess.  Squares are the Sgr dSph stars;
triangles are the two stars in Fornax and the star symbol represents the Galactic
halo carbon star. The lower panel shows the [9]$-$[11] colour, where the
filled squares show Sgr dSph stars with [9]$<6$ mag and the open squares the 
fainter stars.
}
\end{figure}

Fig. \ref{jk} shows the relation between the $J-K$ colour and the $K-[9]$ and
$K-[11]$ colours. Stars in Sgr dSph  are shown corrected for interstellar reddening
(using the values listed in section 2.2, with a higher value for Sgr01).
Stars with $J-K<2$ exhibit little reddening and on average only a small
mid-infrared excess. 
Among the
blue stars, only Sgr13 shows significant excess flux at 9 $\mu$m.

For redder stars,  the $J-K$ colour and the mid-infrared excess show
correlated evidence for circumstellar dust. 

The lower panel shows the [9]$-$[11] colour as function of $J-K$.  The
observations show no clear trend, although a trend does appear when the
stars are separated on the basis of their 9$\mu$m magnitude. The open squares
show the Sgr dSph stars with [9]$>6$ and the closed squares those with [9]$<6$. The
brighter stars generally have redder $J-K$ colours. This can be understood
as more dust (brighter [9]) resulting in higher optical depths and thus more
reddening at $J-K$. 

\subsection{Bolometric magnitudes}
\label{Bol_mag}

The  $JHK$ magnitudes and the bolometric corrections derived by
Whitelock et al. (2006) were used to estimate the apparent bolometric
magnitudes of the programme stars:
\begin{eqnarray}
\label{bol_cor}
\nonumber {\rm BC_K} & = & +\, 0.972 + 2.9292\times(J-K)
  -1.1144\times(J-K)^2 \\
 & & +0.1595\times(J-K)^3 -9.5689\,10^{-3}(J-K)^4.
\end{eqnarray}
\noindent 
Because this equation is based on magnitudes from the SAAO system, we converted all
the photometry to the SAAO system and then reddening corrected it.  The absolute bolometric
magnitudes were then calculated assuming distance moduli of 17.02 for Sgr dSph
(Mateo et al., 1995) and 20.66 for Fornax (Bersier 2000).  The results for
the Sgr dSph stars are listed in Table \ref{photo}.

\begin{figure}
\includegraphics[width=\columnwidth,clip=true]{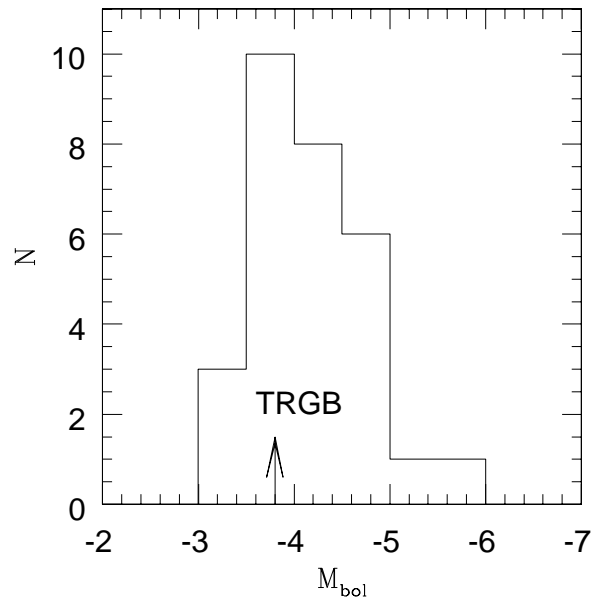}
\caption{\label{histo} Distribution of bolometric magnitudes of the observed
stars in Sgr dSph, assuming a distance modulus of 17.02. }  
\end{figure}

The histogram of the luminosity distribution is shown in Fig.~\ref{histo}.
The tip of the RGB (TRGB) is also indicated, using the calibration of
Bellazzini et al. (2001). The large majority of stars are within one magnitude
of this tip. One star (Sgr27) is significantly below the TRGB (by $\sim$ 0.7 mag).

\begin{table*}
\caption[]{\label{log_obs} Log of the observations. Names and  adopted 
coordinates. $t_{\rm PAH1}$ and $t_{\rm PAH2}$ are the exposure time using 
the filters centred at 8.59 and 11.25$\mu$m respectively. Stars with names
 beginning with WMIF in column 3 are taken from the sample of Whitelock et al. (1999).}
\begin{center}
\begin{tabular}{lllllccl}
\hline
Star number &  2MASS name & Previous name &RA & Dec  & t PAH1 & t PAH2 &Ref star   \\
 & & &\multicolumn{2}{c}{(J2000)}  & s& s  \\
\hline
\textbf{Sgr dSph} &       &    & && &  &  \\
Sgr01&  18405750$-$2734227& &18 40 57.5 & $-$27 34 22.7&90.16 &90.62& HD119193 \\
Sgr02&  18414350$-$3307166& IRAS 18384$-$3310  &18 41 43.50 & $-$33 07 16.6& 180.32&181.24&HD169916\\
Sgr03&  18443095$-$3037098& IRAS 18413$-$3040	 &18 44 30.96 & $-$30 37 09.8& 180.32&181.24&HD169916\\
Sgr04&  18450563$-$2915574& &18 45 05.64 & $-$29 15 57.4& 90.16&90.62&HD169916\\
Sgr05&  18463228$-$2953484& &18 46 32.3 & $-$29 53 48.4 &180.32 & 59.1& HD169916 \\
Sgr06&  18463912$-$3045527& WMIF-C02 &18 46 39.12 & $-$30 45 52.7 &1081.92 &1449.92 &HD169916 \\
Sgr07&  18465160$-$2845489& IRAS 18436$-$2849	  &18 46 51.6 & $-$28 45 48.9 &90.16 &90.62& HD181109 \\
Sgr08&  18511108$-$3121563& &18 51 11.08 & $-$31 21 56.3 &360.64 &362.48&HD169916 \\
Sgr09&  18514105$-$3003377& &18 51 41.05 & $-$30 03 37.7 & 90.16&90.62& HD177716 \\
Sgr10&  18524416$-$3121150& &18 52 44.16 & $-$31 21 15.0  &90.16 &90.62&HD169916 \\
Sgr11&  18525030$-$2956317& WMIF-C04  & 18 52 50.7 & $-$29 56 30.6 &540.96 &634.34& HD181109 \\
Sgr12&  18534097$-$2934228& &18 53 40.98 & $-$29 34 22.9 & 90.16&90.62& HD169916 \\
Sgr13&  18542429$-$3025105& WMIF-C06&18 54 24.29 & $-$30 25 10.6 &721.28&  226.55  &HD177716\\
Sgr14&  18532937$-$2938241& WMIF-C05&18 53 29.38& $-$29 38 24.2 &1803.2 & &HD169916 \\
Sgr15&  18584385$-$2956551& IRAS 18555$-$3001	&18 58 43.85  & $-$29 56 55.1 &90.16 &90.62& HD181109\\
Sgr16&  19000215$-$3035347& IRAS 18568$-$3039 &19 00 02.15  & $-$30 35 34.7 &90.16 &90.62&HD181109 \\
Sgr17&  19015287$-$3032391& WMIF-C15&19 01 52.88 & $-$30 32 39.2 &270.48 &362.48&HD169926 \\
Sgr18&  19043562$-$3112564& IRAS 19013$-$3117 &19 04 35.62 & $-$31 12 56.4 &90.16 &90.62&HD181109 \\
Sgr19&  19044898$-$3110540& WMIF-C17&19 04 48.98 & $-$31 10 54.0 &360.64& 362.48 & HD169916\\
Sgr20&  19065718$-$3412371& &19 06 57.18 & $-$34 12 37.1 &450.80 &724.96&HD181109 \\
Sgr21&  19093902$-$2956561& WMIF-C18&19 09 39.03 & $-$29 56 56.1 & 721.28&634.340 &HD169916 \\
Sgr22&  19103987$-$3228373& IRAS 19074$-$3233 &19 10 39.87 & $-$32 28 37.3 &90.16 &90.62&HD177716 \\
Sgr23&  19125006$-$3051228& &19 12 50.06 & $-$30 51 22.8 &90.16 &90.62&HD177716  \\
Sgr24&  19165541$-$3248289& &19 16 55.41 & $-$32 48 28.9 &180.32 &453.1&HD181109 \\
Sgr25&  19232461$-$3400149& WMIF-C22&19 23 24.61 & $-$34 00 15.0&1803.2&    &HD169916 \\
Sgr26&  19234548$-$3431150& WMIF-C23  &19 23 24.48 & $-$34 31 15.0 &1803.2 & &HD169916 \\
Sgr27&  19313851$-$3002305& WMIF-C25 &19 31 38.5 & $-$30 02 30.5  &1803.2& &HD177716\\
Sgr28&  19425082$-$3334333& WMIF-C26&19 42 50.83 & $-$33 34 33.3&721.28&  &HD177716\\
Sgr29&  19485065$-$3058317& &19 48 50.65 & $-$30 58 31.9& 90.16&90.62&HD181109\\
\hline
\textbf{Fornax} &           & && &  &  \\
For1&   02385056$-$3440319& &02 38 50.56 & $-$34 40 31.9 &2163.84 & &HD16815 \\
For2&   02391232$-$3432450& &02 39 12.33 & $-$34 32 45.0 & 1803.20&&HD16815 \\
\hline
\textbf{Halo} &           && && &  &  \\
Halo1&  20202766$-$1449272& IRAS 20176-1458&20 20 27.66 & $-$14 49 27.1 &90.16 &90.62&HD177716 \\
\hline \\
\end{tabular}
\end{center}
\end{table*}

\begin{table*}

\caption[]{\label{photo} Sgr dSph and Fornax  photometry.
For stars with names beginning with WMIF (Whitelock et al. 1999)) in column 3 of Table \ref{log_obs}, the $JHK$
photometry comes from unpublished data taken at SAAO on the 1.9m telescope (SAAO in last column),
and, where multiple observations are available, refers to the mean
magnitude. For all other stars (2MASS in last column), the $JHK_s$ is taken from 2MASS. All $JHK_s$
magnitudes are given on the 2MASS photometric system.  PAH1 and PAH2
represent our VISIR observations at 8.59 and 11.25$\mu$m respectively.
Zero-points are taken to be  53.7 (8.59$\mu$m), 31.5 (11.59$\mu$m) and 28.3 Jy
(IRAS).}
\begin{center}
\begin{tabular}{lllllllll}
\hline
Star number & $J$ & $H$ & $K_s$ & PAH1&PAH2& M$_{bol}$& IRAS [12]  & $J$$H$$K_s$ observations\\
  & mag & mag & mag &mag&mag &mag &mag& \\
\hline
\textbf{Sgr D} &           & && &  & & &\\
Sgr01&12.479 &10.774  &9.308 &5.48$\pm$0.03 &5.11$\pm$0.03 & $-$4.66 & &2MASS\\
Sgr02&11.577 &9.971   &8.575 &4.96$\pm$0.02 &4.24$\pm$0.03 &$-$5.29 &3.90&2MASS\\
Sgr03&13.236 & 11.183 &9.612 &6.03$\pm$0.04 &5.62$\pm$0.06 &$-$4.67 &5.50&2MASS\\
Sgr04&12.250 &10.708  &9.438 &7.38$\pm$0.12 &7.08$\pm$0.24 &$-$4.33 &&2MASS\\
Sgr05&13.027 &11.319  &9.866 &6.73$\pm$0.05 &6.41$\pm$0.16 & $-$4.09 &&2MASS\\
Sgr06&11.481 &10.315  &9.604 &8.80$\pm$0.08 &7.75$\pm$0.09 &$-$3.99  &&SAAO\\
Sgr07&13.060 &11.123  &9.599 &5.86$\pm$0.04 &5.18$\pm$0.05 & $-$4.56 &4.84&2MASS\\
Sgr08&13.062 &11.528  &10.320 &8.29$\pm$0.14 &7.76$\pm$0.22& $-$3.42 &&2MASS \\
Sgr09&12.090 &10.570  &9.420  &7.00$\pm$0.09 &6.63$\pm$0.15 &$-$4.29 &&2MASS \\
Sgr10&12.045 &10.367  &9.048 &6.37$\pm$0.04 &6.13$\pm$0.10 &$-$4.81  &&2MASS\\
Sgr11&11.219 &10.058   &9.448 &8.29$\pm$0.11 &7.71$\pm$0.13 & $-$4.16 &&SAAO\\
Sgr12&13.258 &11.599  &10.182&6.78$\pm$0.07 &6.84$\pm$0.20 &$-$3.73  && 2MASS\\
Sgr13&11.424 &10.298  &9.724 &8.78$\pm$0.08 &     &$-$3.91 &&SAAO\\
Sgr14&11.445 &10.414  &9.904 &9.42$\pm$0.11 &     &$-$3.80  &&SAAO\\
Sgr15&13.363 &11.331  &9.701 &6.01$\pm$0.04 &5.41$\pm$0.06 &$-$4.61 &5.28&2MASS\\
Sgr16&14.644 &12.555  &10.685&5.70$\pm$0.03 &5.32$\pm$0.07 &$-$3.87 &5.55&2MASS\\
Sgr17&11.874 &10.511  &9.526 &7.87$\pm$0.06 &      &$-$4.08 & &SAAO\\
Sgr18&14.227 &11.956  &10.168&5.37$\pm$0.03 &4.80$\pm$0.04 &$-$4.46 &4.63 &2MASS\\
Sgr19&12.603 &11.078  &9.967 &7.59$\pm$0.08 &7.05$\pm$0.11 &$-$3.73 & &SAAO\\
Sgr20&13.112 &11.533  &10.289&7.65$\pm$0.07 &6.35$\pm$0.07 &$-$3.48 &&2MASS \\
Sgr21&10.862 &9.751   &9.099 &7.67$\pm$0.07 &7.87$\pm$0.14 &$-$4.52 &&SAAO\\
Sgr22&11.957 &9.991   &8.349 &4.29$\pm$0.01 &3.36$\pm$0.01 &$-$5.92 &3.03&2MASS\\
Sgr23&12.743 &11.084  &9.736 &6.76$\pm$0.07 &5.75$\pm$0.06 &$-$4.13 &&2MASS\\
Sgr24&12.118 &10.643  &9.513 &7.65$\pm$0.11 &6.98$\pm$0.09 &$-$4.17 &&2MASS\\
Sgr25&11.209 &10.235  &9.793 &8.00$\pm$0.04&     &$-$3.98 &&SAAO\\
Sgr26&11.303 &10.359  &9.882 &8.82$\pm$0.09&     &$-$3.89 & &SAAO\\
Sgr27&12.347 &11.192  &10.473&9.05$\pm$0.13 &     &$-$3.12  &&SAAO\\
Sgr28&11.424 &10.401  &9.961 &8.42$\pm$0.18 &      &$-$3.78 &&SAAO\\
Sgr29&13.599 &11.741  &10.209&6.48$\pm$0.07 &6.38$\pm$0.12 &$-$3.90  &&2MASS\\
\hline
\textbf{Fornax} &           & && &  &  & \\
For1&14.722 &13.262  &12.120&9.53$\pm$0.13 &     &$-$5.18 &&2MASS\\
For2&  16.106&   14.525& 12.879&9.56$\pm$0.20&     &$-$4.76 &&2MASS\\
\hline
\textbf{Halo} &           & && &  &  & \\
Halo1&11.850 &10.143  &8.707 &5.50$\pm$0.03 &4.74$\pm$0.04 &N/A  &4.50&2MASS\\

\hline \\
\end{tabular}
\end{center}
\end{table*}

\section{Methods to determine mass-loss rates from infrared colours}
\label{other_methods}

The mid-infrared colours can be used to obtain mass-loss rates from these
AGB stars. From a survey of Miras in the South Galactic Cap, Whitelock et
al. (1994) showed that the mass-loss rates and the $K-[12]$ colour are
tightly correlated, where [12] is the IRAS 12$\mu$m magnitude (see fig.\ 21
in their paper). This can be understood from the fact that the $K$ magnitude
is a measure of the emission from the star, while the [12] magnitude a
measure of emission from dust in the circumstellar envelope. The $K-[12]$
colour will thus increase with mass-loss rate.

Mass-loss rates from Galactic AGB stars have also been estimated using dust
radiative transfer models (e.g., Le Bertre 1997). For these models, a dust
composition is assumed, as well as a dust grain size, a density distribution,
an expansion velocity for the envelope, and a temperature and luminosity for the
central star.  The optical depth is a free parameter. The best
fit model of the observed SED gives an estimate of the optical depth of the
dusty envelope. If we assume that dust is composed of spherical grains of a
given composition and size, we can determine their optical properties using
Mie theory. The expansion velocity of circumstellar envelopes of Galactic AGB
stars have been estimated using CO observations (e.g., Loup et al. 1993). Thus
by determining the optical depths of the envelopes, we can estimate mass-loss
rates. The values estimated with these dusty radiative transfer models
are consistent with those derived from the CO emission line measurements
(e.g., Le Bertre 1997). For more information about infrared methods to determine
 mass-loss rates from AGB stars, we refer to the review by van Loon (2006).

\section{Methods for determining the mass-loss rate}
\subsection{Method 1: calibrated colour relations}

\subsubsection{Calibration}

As mentioned in Section~\ref{other_methods}, a tight correlation exists 
between the $K-[12]$ colour and the mass-loss rates for Galactic
AGB stars (Whitelock et al. 1994). Here, the 12$\mu$m magnitude represents the
broad-band IRAS measurement. The Whitelock et al.  sample contains 58 O-rich
stars and only 3 C-rich stars. Using a further sample of 239 Galactic C-rich
stars, Whitelock et al. (2006) determined the correlation between $K-[12]$ and
mass-loss rates for C-rich AGB stars. This relation was quantified as:
\begin{eqnarray}
\label{mass-loss}
\nonumber \log(\dot{M}_{\rm total}) &&=-7.668+0.7305(K-[12])\\
\nonumber  &&-5.398 \times10^{-2}(K-[12])^2  \\
  &&+1.343\times10^{-3}(K-[12])^3
\end{eqnarray}

Our observations used VISIR filters with wavelengths close to 12 $\mu$m. The
two filters, selected because of their high sensitivity, have
wavelengths centred at 8.59 and 11.25$\mu$m. To derive an
equivalent relation to Eq. \ref{mass-loss} for these filters, we retrieved
all the available IRAS LRS spectra used by Whitelock et al. (2006). We then
convolved these spectra with the VISIR transmission curves for the filters
used. We also convolved them with the IRAS 12$\mu$m filter, and scaled the
spectra such that the catalogued IRAS 12$\mu$m flux was recovered.  The [9]
and [11] magnitudes of the Galactic C-stars were obtained using the
zero-points listed in Table \ref{photo}.

The $K$ band flux of the stars we observed in Sgr dSph and Fornax have been
measured either with 2MASS ($K_{\rm S2MASS}$) or at the SAAO ($K_{\rm
SAAO}$). To obtain a uniform flux measurement for the $K$-band we
transformed the SAAO measurements to the 2MASS photometric system, following
(Carpenter 2001):

\begin{equation}
K_{\rm S2MASS}=K_{\rm SAAO}+0.020(J-K)_{\rm SAAO}-0.025
\end{equation} 

This was done for all stars in the Whitelock et al. (1994, 2006) samples.  The
resulting colours were plotted against the dust mass-loss rate taken from
the Whitelock et al. papers. 

\begin{figure*}
\includegraphics[width=\textwidth,clip=true]{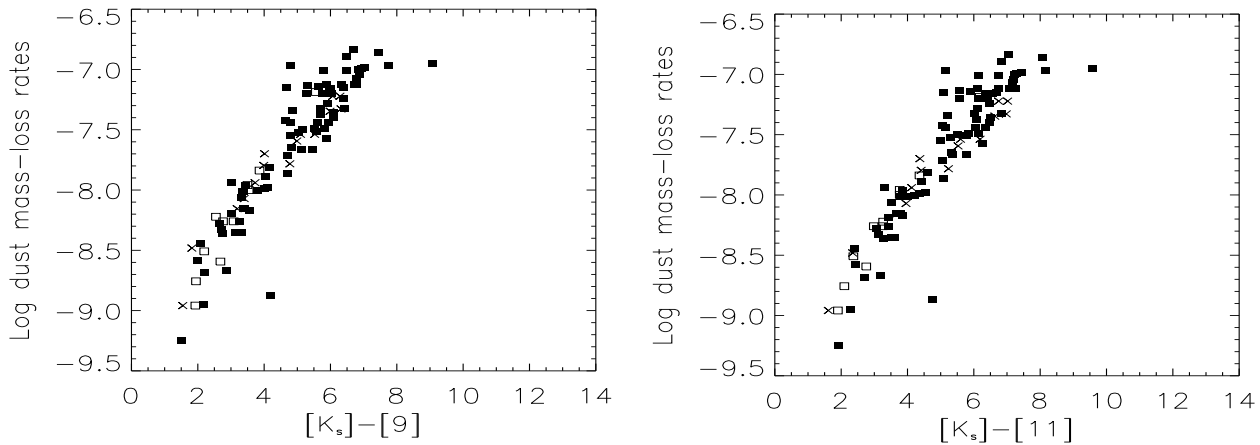}
\caption{\label{massloss_all_gal} Dust mass-loss rates of C-rich AGB stars
as a function of $K-[9]$ (left) and $K-[11]$ (right) colours of stars from the
sample in Whitelock et al. (2006) (Galaxy, black squares), Lagadec et
al. (2007) (SMC, open squares) and Zijlstra et al. (2006) (LMC, crosses).}
\end{figure*}

The result for the carbon-rich sample of Whitelock et al. (2006) is shown in
Fig. \ref{massloss_all_gal}: there is a good correlation between the
$K_s-[9]$
and $K_s-[11]$ colours and the dust mass-loss rates. The oxygen-rich stars
of Whitelock et al. (1994) show a similar relation (Fig.
\ref{massloss_all_gal}) but extending to less red colours. The relation
holds for $K_s-[9]<7$ and is a consequence of dust emission at
10$\mu$m and dust opacity in the $K_s$-band.

Note that Whitelock et al.  derived total mass-loss rates (dust+gas), adopting
a gas-to-dust mass ratio of 200. As what we observed at 9 and 11$\mu$m is
mostly due to emission from dust, we determine dust mass-loss
rates. The gas-to-dust mass ratios in the envelope of AGB stars in the
observed galaxies being poorly known, we prefer to discuss dust mass-loss
rates.

The relations between colours and mass-loss rates for these Galactic
AGB carbon stars can be quantified as:
\begin{eqnarray}
\label{mass-loss9}
\nonumber \log(\dot{\rm M}_{\rm dust}) &=& -9.41+0.27(K_s-[9])+0.05(K_s-[9])^2\\
   & & -0.0055(K_s-[9])^3
\end{eqnarray}
and:
\begin{eqnarray}
\label{mass-loss11}
\nonumber \log(\dot{\rm M}_{\rm dust}) &=& -9.58+0.26 (K_s-[11])
      +0.05(K_s-[11])^2\\
      & & -0.0053(K_s-[11])^3
\end{eqnarray}

To check the dependence of this relation on metallicity, we determined the
equivalent relations for stars in the Small and Large Magellanic Clouds
(hereinafter SMC and LMC), which we assume to have lower metallicity than
similar stars in the Milky Way. Recent Spitzer observations have yielded
reliable determinations of mass-loss rates from AGB stars in these
galaxies. We used the sample of carbon stars in the SMC presented by Lagadec
et al. (2007), and in the LMC by Zijlstra et al. (2006). The mass-loss rates
of these stars have been determined by Groenewegen et al. (2007) using a
radiative transfer model. They give total mass-loss rates, using a
gas-to-dust mass ratio of 200 and assuming an expansion velocity for all the
stars in the sample of 10\,km\,s$^{-1}$.

We obtained the equivalent [9] and [11] magnitudes for the stars in the SMC
and LMC sample by convolving the Spitzer IRS spectra with the response
curves of the VISIR filters. The $K$ magnitudes were taken from the
Groenewegen et al. (2007) tabulation.  The resulting relations for $K_s-[9]$
and $K_s-[11]$ versus mass-loss rate are shown in Fig.
\ref{massloss_all_gal}. The AGB stars in the Galaxy, SMC and LMC all follow
the same trend on a $K_s -[9]$ or $K_s -[11]$ versus dust mass-loss rate
relation. This appears not to depend on metallicity as what we measure is an optical depth and can be applied to
derive the dust mass-loss rates for AGB stars in other galaxies.

\begin{figure}
\includegraphics[width=\columnwidth,clip=true]{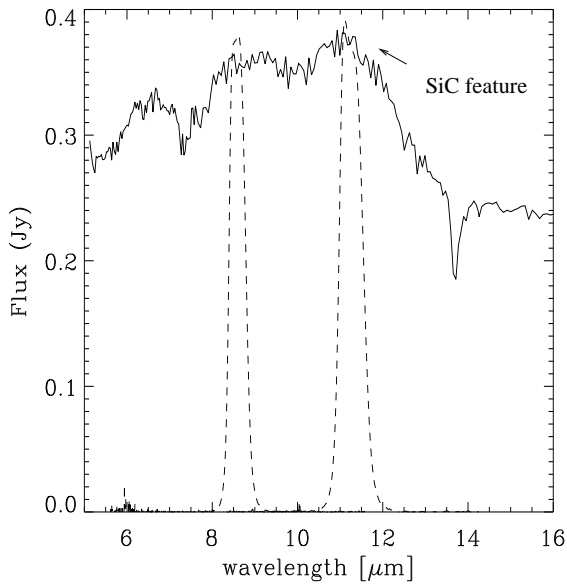}
\caption{\label{spitzer}Spitzer spectrum of the C-rich SMC AGB star IRAS 00554 
(Lagadec et al. 2007), superimposed with the VISIR PAH1 and PAH2 filters.}  
\end{figure}

The Spitzer spectra of LMC and SMC carbon stars show the presence of deep molecular absorption bands,
together with the 11.3$\mu$m SiC emission band, making it difficult to
define the continuum (Fig.~\ref{spitzer}). Zijlstra et al. (2006) define
four narrow bands suitable for measuring the continuum for carbon stars. One
of these (the 9.3$\mu$m band) is close to the transmission of the VISIR
9$\mu$m filter. This filter is in a region relatively free from strong molecular
bands, apart from a possible C$_3$ band (Zijlstra et al. 2006), providing a metallicity-free
mass-loss estimation. The second filter coincides with the 11.3$\mu$m band. The
SiC band is known to be metallicity dependent, being weaker for lower
metallicity stars (Lagadec et al. 2007).  The $K_s-[11]$ relation is therefore
expected to have some metallicity dependence, but the SiC band does not
dominate the spectrum in the same way as the silicate band does in
oxygen-rich stars, and the dependence is therefore limited.

\subsubsection{Application}

The $K_s-[9]$ and $K_s-[11]$ colours thus lead immediately to an estimation of
the dust mass-loss rates in Sgr dSph and Fornax.


In this work, we assume that all the AGB stars observed are C-rich. This is
likely given the low metallicities of Sgr dSph and Fornax, although we cannot
rule out that a few stars in our sample might be O-rich. However, the
relation between $K_s-[9]$ or $K_s-[11]$ and mass-loss rates are similar for
C-rich and O-rich stars. We should keep in mind that the methods used
to determine mass-loss rates in the Galaxy and the Magellanic Clouds are not
direct measurements. The mass-loss rates in the Galaxy were calculated using
the Jura (1987) formalism, which allows one to estimate the mass-loss rates
for carbon stars using the observed IRAS 60$\mu$m emission and measured
expansion velocities. The mass-loss rates in the Magellanic Clouds have been
determined using a radiative transfer model to fit mid-infrared spectra and
near-infrared photometry, assuming a constant outflow velocity of 10 $km$.$s^{-1}$.
 These are the probably the most reliable estimates
available. The [9] and [11] magnitudes are calculated from our VISIR
observations and the derived dust mass-loss rates are listed
in Table~\ref{mass_loss}.
 
\subsection{Method 2: radiative transfer models}

To provide an alternative estimate for the mass-loss rates for these AGB
stars, we also computed a grid of dust radiative transfer models using the
DUSTY 1D code (Ivezi\'c et al. 1999). For each model, the density
distribution of a radiation-driven wind was assumed, $\rho\sim r^{-1.8}$
(Elitzur \& Ivezi\'c 2001). For this choice, the DUSTY models compute the
wind structure by solving the hydrodynamical equations as a set, coupled to
the radiative transfer.

We assumed that the central star emits like a blackbody with T=2800K, a
typical temperature for an AGB star.  Previous papers have shown that in low
metallicity galaxies, most of the mass-losing AGB stars are carbon-rich (e.g. van
Loon et al. 1999a, Matsuura et al. (2002, 2005)). The dust in the stars we
observed is therefore likely to be carbonaceous and this is what we assume. 
Specifically, we assume that the dust grains are spherical and composed of a
mixture of amorphous carbon (95\%) and SiC (5\%), with a grain size of
0.1$\mu$m. The optical properties of these grains are calculated using
Hanner (1988) and P\'egouri\'e (1988) respectively. We assumed that the dust
temperature at the inner boundary is 1000K. The shell thickness (i.e. the
ratio between the outer an inner radius) is set to 10$^4$.

The optical depth of the envelope is proportional to the dust mass-loss
rate. We can thus associate a mass-loss rate with a model of a given optical
depth. The output of these DUSTY models contains values of the total
mass-loss rate (dust+gas) and of the shell expansion velocity. As stated
above, we prefer to discuss dust mass-loss rates. We convert the total
mass-loss rates to dust-mass loss rates by dividing by the gas-to-dust mass
ratio. We, however, have to keep in mind that the expansion velocity also
depends on the dust-to-gas ratio, rendering the total mass-loss rate less than
straightforward to measure.

\begin{figure}
\includegraphics[width=\columnwidth,clip=true]{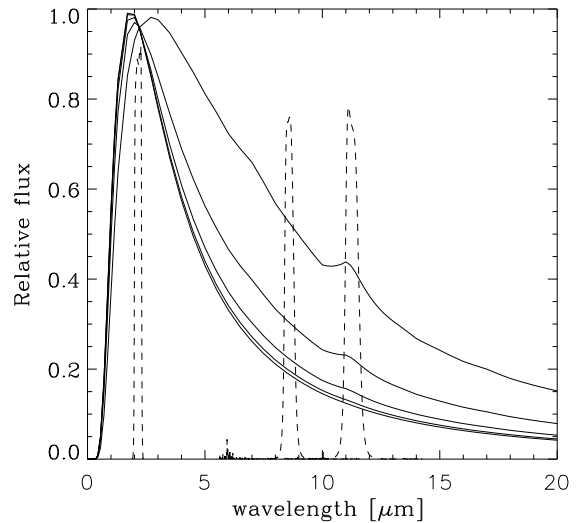}
\caption{\label{dusty_filters} Spectral energy distribution of DUSTY models pf carbon dust
with different optical depths (respectively 0.01, 0.3, 1., 10. and 31. at 1$\mu$m),
 superimposed with the 2MASS $K_s$ and VISIR PAH1 and 
PAH2 filters.}  
\end{figure}

We thus calculated a benchmark of DUSTY models with these input parameters and
different optical depths.  The outputs of these models are spectra which we
convolved with the 2MASS $K_s$ and VISIR PAH1 and PAH2 filters
(Fig. \ref{dusty_filters}) to determine the $K_s-[9]$ and $K_s-[12]$ colours
associated with the different models.

Using these models, we find these relations between colours and dust mass-loss
rates:

\begin{equation}
\label{mass3}
\log(\dot{\rm M}_{\rm dust})=-8.14 \times ((K_s-[9])-0.97)^{-0.06}\times 0.75 \log(L/10^4)
\end{equation}

\begin{equation}
\label{mass4}
\log(\dot{\rm M}_{\rm dust})=-8.28 \times ((K_s-[11])-0.98)^{-0.07}\times 0.75 \log(L/10^4)
\end{equation}
where $L$ is the luminosity of the star in units of solar luminosity.
The measured colours then directly give us an estimation of the mass-loss
rates using relation (\ref{mass3}) and (\ref{mass4}). The resulting values 
are listed Table \ref{mass_loss}.

\subsection{Comparison}

\begin{figure*}
\includegraphics[width=16cm,clip=true]{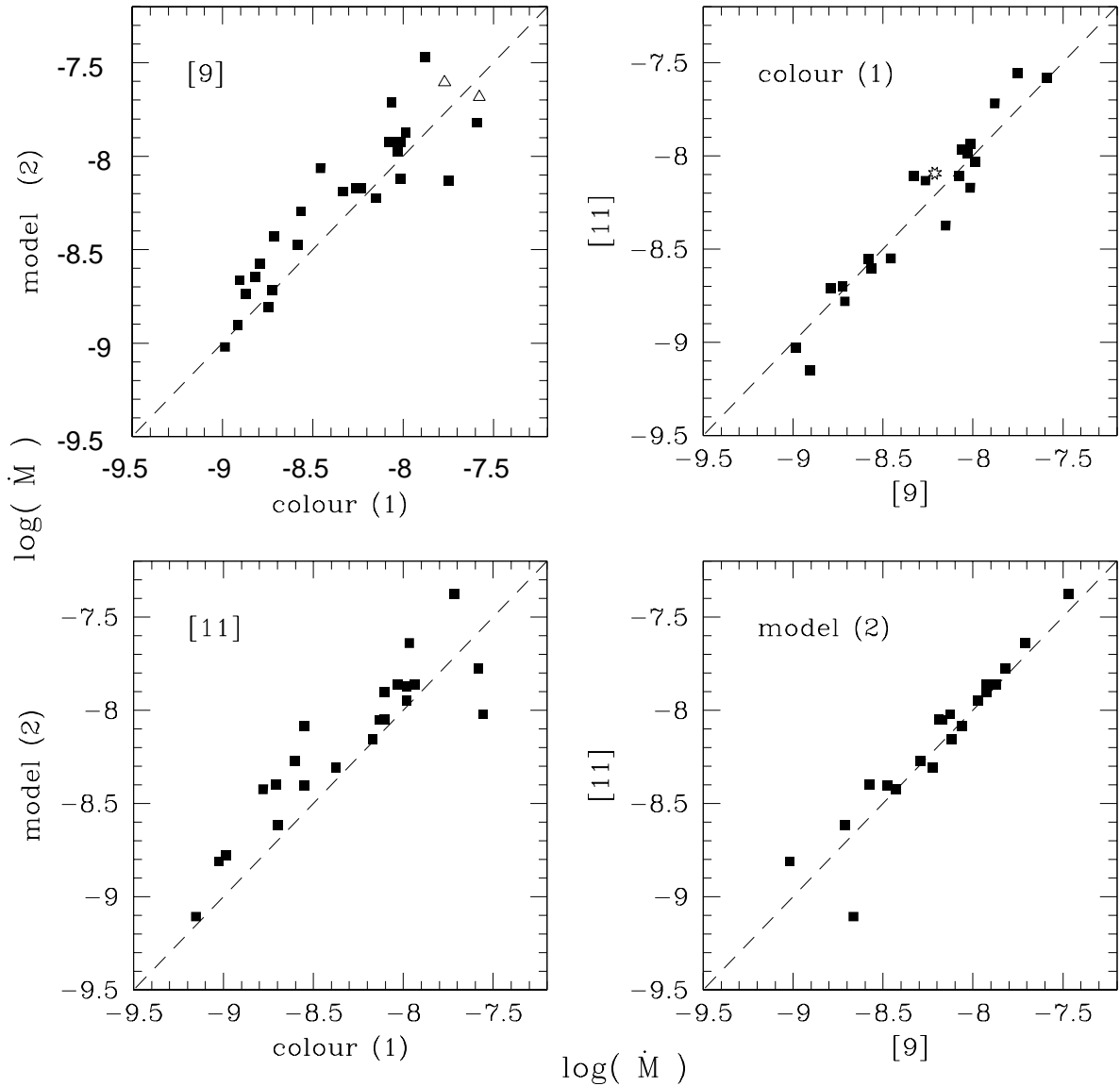}
\caption{\label{comp} Comparison between the different determinations
of the dust mass-loss rate. The left panels plot the rates determined from the
colour method (method 1) against those from the DUSTY models (method 2),
for the two VISIR filters. The right panels compare the rates derived from
the two different filters, for each of the methods. Squares are Sgr dSph stars,
triangles Fornax and the star symbol is the Galactic halo star.}  
\end{figure*}

Both methods should only be used for the colour range for which they are
defined. This is especially true for blue stars, where the mass-loss rates
do not go to zero for photospheric colours (as it is a logarithmic equation).
We choose to set mass-loss rates for stars with $K_s-[9]<0.8$ to zero.
This affects the three bluest stars: Sgr06, Sgr13 and Sgr14.

Variability introduces an uncertainty. For the stars with WMIF names, the
$K$-band magnitudes are averaged over the variability. For the others, they
are single-epoch values which may differ from the mean: the amplitude at $K$
is typically 0.1 mag for blue stars and can be up to 1 mag or possibly more
for red stars. Finally, the VISIR photometry may also be affected by stellar
variability. Indeed little is known about the variability in the mid-infrared
 of carbon AGB stars. Observations of a sample of 23 C-rich AGB stars (Le Bertre 1992)
 have shown that the amplitude of variation of the  observed stars from 1 to 20$\mu$m
 seems to decrease with wavelength.  Both effects (near-infrared and mid-infrared variability)
 will affect all four methods in the same way.

The different methods also show internal differences.   Fig.
\ref{comp} shows the comparison. The left panels compare mass-loss rates
derived from the colours (method 1) with those from the models (method 2).
A systematic offset can be seen, with the DUSTY models giving a higher
mass-loss rate by up to a factor of 1.5. The right panels compare
the mass-loss rates derived from the two different filters used, where the
agreement is good.  There is a small offset seen in the model results,
where the mass-loss rates derived using [11] are a little higher than those
derived from [9]. The offset amounts to about 10 per cent. This may
be due to a contribution from SiC at 11$\mu$m. 

The first aim of our DUSTY model was to check the validity of our mass-loss rates
 determination method based on infrared colours. The fact that both methods give 
very similar results thus indicates that this method is a good estimator of AGB stars dust-mass loss rates.

\section{Discussion}

\subsection{ Mass-loss rates}

The first main result of this study is that most of the observed stars in
Sgr dSph (29 stars) and Fornax (2 stars) are red in $K_s-[9]$, indicating the
presence of a dusty envelope and dusty mass loss. The dust mass-loss rates
are in the range 5$\times10^{-10}$ to $3\times10^{-8}$ M$_{\odot}$yr$^{-1}$
for the AGB stars in Sgr dSph and around 5$\times10^{-9}$ M$_{\odot}$yr$^{-1}$
for those in Fornax. A number of Sgr dSph stars have $K_s-[9]<1$, and these are
probably photospheric colours.

 Our observations indicate that even at very low metallicities ([${\rm Fe}/ {\rm H}$]$\sim$-0.9, see Section \ref{tar_sel}), as found in
the Fornax dwarf spheroidal galaxy, dusty mass-loss occurs.
\begin{table*}
\caption[]{\label{mass_loss} Mass-loss rates determined from our VISIR observations. $\dot{M}_1$ and$\dot{M}_2$ are the dust mass-loss rates (in M$_{\odot}$yr$^{-1}$) derived from method 1, using the [9] and [11] colours respectively. $\dot{M}_3$ and $\dot{M}_4$ are the dust mass-loss rates (in M$_{\odot}$yr$^{-1}$) derived from method 2, using the [9] and [11] colours respectively. }
\begin{center}
\begin{tabular}{lllllll}
 \hline Star number &K$_s$--[9]&K$_s$--[11]&$\dot{M}_1$& $\dot{M}_2$ &$\dot{M}_3$ &$\dot{M}_4$
\\

\hline
\textbf{Sgr dSph} &           & && &  &  \\
\hline
Sgr01&3.74  &4.11   &1.0$\times10^{-8}$ & 9.2$\times10^{-9}$ &  1.4$\times10^{-8}$& 1.5$\times10^{-8}$ \\
Sgr02&3.56  &4.28   &8.6$\times10^{-9}$&  1.1$\times10^{-8}$ &  2.1$\times10^{-8}$& 2.4$\times10^{-8}$\\
Sgr03&3.53  &3.94   &8.4$\times10^{-9}$ & 7.9$\times10^{-9}$&  1.3$\times10^{-8}$&  1.4$\times10^{-8}$\\
Sgr04& 2.00 &2.30  &1.9$\times10^{-9}$ & 1.7$\times10^{-9}$ &3.9$\times10^{-9}$ & 3.9$\times10^{-9}$\\
Sgr05&3.08  &3.40   &5.4$\times10^{-9}$&  4.7$\times10^{-9}$ &7.2$\times10^{-9}$& 7.2$\times10^{-9}$  \\
Sgr06&0.75  & 1.80   &6.6$\times10^{-10}$ & 1.0$\times10^{-9}$&  blue&  1.6$\times10^{-9}$\\
Sgr07&3.68  &4.36   &9.7$\times10^{-9}$&  1.2$\times10^{-8}$ & 1.3$\times10^{-8}$& 1.5$\times10^{-8}$ \\
Sgr08&1.98  &2.51   &1.9$\times10^{-9}$&  2.0$\times10^{-9}$ & 2.0$\times10^{-9}$& 2.5$\times10^{-9}$ \\
Sgr09&2.37  &2.74   &2.7$\times10^{-9}$&  2.5$\times10^{-9}$ & 5.3$\times10^{-9}$& 5.5$\times10^{-9}$ \\
Sgr10&2.62  &2.86   &3.5$\times10^{-9}$&  2.8$\times10^{-9}$ & 9.1$\times10^{-9}$& 8.7$\times10^{-9}$ \\
Sgr11&1.10  &1.68   &8.7$\times10^{-10}$ & 9.4$\times10^{-10}$& 3.0$\times10^{-10}$&  1.5$\times10^{-9}$\\
Sgr12&3.35  &3.29   &7.0$\times10^{-9}$ & 4.2$\times10^{-9}$ & 6.4$\times10^{-9}$& 5.3$\times10^{-9}$ \\
Sgr13& 0.89 &   &7.3$\times10^{-10}$&   & blue& \\  
Sgr14& 0.43&  &  5.2$\times10^{-10}$  & & blue& \\
Sgr15&3.64  &4.24   &9.3$\times10^{-9}$& 1.0$\times10^{-8}$& 1.3$\times10^{-8}$& 1.5$\times10^{-8}$\\
Sgr16&4.93  &5.31   &3.0$\times10^{-8}$&  2.6$\times10^{-8}$&  1.2$\times10^{-8}$&  1.3$\times10^{-8}$\\
Sgr17&1.60  & & 1.3$\times10^{-9}$  && 1.9$\times10^{-9}$ &\\
Sgr18&4.74  &5.31   &2.6$\times10^{-8}$ & 2.6$\times10^{-8}$&  1.7$\times10^{-8}$& 1.9$\times10^{-8}$\\
Sgr19&2.32  &2.86  &2.6$\times10^{-9}$ & 2.8$\times10^{-9}$& 3.5$\times10^{-9}$&  4.1$\times10^{-9}$\\ 
Sgr20&2.58  &3.88   &3.4$\times10^{-9}$ & 7.5$\times10^{-9}$ &  3.5$\times10^{-9}$& 6.0$\times10^{-9}$\\
Sgr21& 1.37 &1.17   &1.1$\times10^{-9}$ & 6.1$\times10^{-10}$ &1.5$\times10^{-9}$ & 3.0$\times10^{-10}$ \\
Sgr22& 4.00 &4.93   &1.3$\times10^{-8}$ & 1.9$\times10^{-8}$ &  3.8$\times10^{-8}$& 4.7$\times10^{-8}$\\
Sgr23&2.92  &3.93   &4.7$\times10^{-9}$ & 7.8$\times10^{-9}$&  6.8$\times10^{-9}$&  9.5$\times10^{-9}$\\
Sgr24&1.81  & 2.48 &1.6$\times10^{-9}$ & 1.9$\times10^{-9}$&  2.8$\times10^{-9}$&  4.1$\times10^{-9}$\\
Sgr25&1.74  &   & 1.5$\times10^{-9}$&   &  2.2$\times10^{-9}$& \\ 
Sgr26&1.01  &&  8.1$\times10^{-10}$  &4.6$\times10^{-11}$&  &\\
Sgr27& 1.36 &  & 1.1$\times10^{-9}$  && 5.6$\times10^{-10}$ &\\
Sgr28& 1.48&&    1.2$\times10^{-9}$  && 1.2$\times10^{-9}$ &\\ 
Sgr29&3.67  &3.77   &9.6$\times10^{-9}$ & 6.7$\times10^{-9}$&  8.3$\times10^{-9}$&  7.6$\times10^{-9}$\\

\hline
\textbf{Fornax} &           & && &  &  \\
\hline
For1& 2.59 &   &3.4$\times10^{-9}$ & & 1.1$\times10^{-8}$ &\\

For2& 3.32 &   &6.9$\times10^{-9}$ & & 1.2$\times10^{-8}$ &\\

\hline
\textbf{Halo} &           & && &  &  \\
\hline
Halo1&3.21  &3.97   &6.1$\times10^{-9}$ & 8.1$\times10^{-9}$&  N/A& N/A\\

\hline \\
\end{tabular}
\end{center}
\end{table*}

\begin{figure}
\includegraphics[width=\columnwidth,clip=true]{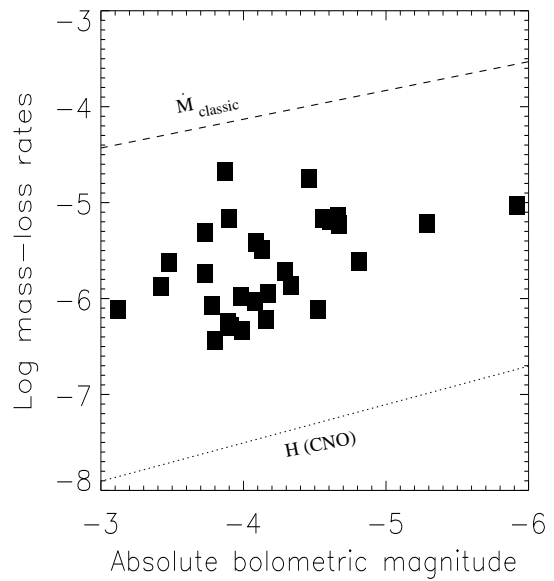}
\caption{\label{mass_loss_lumi} Total (gas+dust) mass-loss rate as a function of the absolute bolometric magnitude.}  
\end{figure}

The mass-loss rates we have derived are \textit{dust} mass-loss rates. To give
constraints to stellar evolution codes, one needs to know the total (gas+dust)
mass-loss rate. For that we need to know the dust-to-gas mass ratio
$\psi$, which is not well known even for Galactic AGB stars. For example, for
IRC\,+10216, the best studied AGB star, values in the range 170-700 are found
in the literature (Mensh'chikov et al. 2001). It is obviously difficult to
determine this ratio in other galaxies. van Loon (2006) argue
that the ratio  scales linearly with the metallicity. This can be simply
understood by the fact that at low metallicity, fewer seeds for dust formation
are available and thus dust formation is less efficient. They therefore 
assume that:

\begin{equation}
\psi=\psi_{\odot}10^{-[\rm Fe/\rm H]}
\end{equation},
where $\psi_{\odot}=0.005$ (van Loon et al. 2005).

If we assume $[$Fe$/$H$]=-0.55$ for Sgr dSph (Dudziak et al. 2000), then for the AGB
stars in this galaxy, $\psi \sim 1.4 \times 10^{-3}$. Assuming $[$Fe$/$H$]=-0.9$
for Fornax leads to an estimate of the dust-to-gas mass ratio $\psi \sim 6.3
\times 10^{-4}$. This will provide total mass-loss rates of 700
and 1600 times, respectively,  higher than the dust mass-loss rates in Sgr dSph
and Fornax. Note, however, that these are upper limits, as the dust expansion
velocity is probably smaller in these galaxies, as well as the drift
velocity, but no observations have yet been made that measure the expansion 
velocities of carbon star at low metallicities.

The assumption of a linear relation between dust production and metallicity
can be questioned. For carbon stars, the two main dust components
are amorphous carbon dust (soot) and silicon carbide. The abundance of the
latter is limited by Si which is not produced in AGB stars. However,
amorphous carbon depends only on the carbon abundance, which is strongly
enriched via third dredge-up. The amount of carbon dust may still depend on
the number of available seeds (e.g. TiC) which can introduce a metallicity
dependence, but it is less likely to be a linear one.


\subsection{Evolution}

For a better understanding of the evolution of these AGB stars, it is
interesting to compare the measured mass-loss rates with the rate at which
mass is consumed by nuclear burning $\dot M_{\rm nuc}$. Most of the energy in AGB
stars comes from CNO burning in the H burning shell. This reaction releases
$\sim 6.1 \times 10^{18}$erg$\,$g$^{-1}$. $\dot
M_{\rm nuc}$ is proportional to the luminosity, L. Fig. \ref{mass_loss_lumi}
shows the measured mass-loss rates as a function of the luminosity.
Over-plotted are the $\dot M_{\rm nuc}$ vs $L$ relation and the classical
single-scattering limit $\dot M_{\rm classic}=L($v$_{\rm exp}$c$)^{-1} \propto
L^{0.75}$\textit{vs} $L$. This diagram clearly indicates that all the measured
mass-loss rates are below the classical single-scattering limit. Thus no
multiple scattering of photons by dust in the envelope is needed to explain
the observed mass-loss rates as can be necessary for the very high mass-loss
rates measured for some Galactic AGB stars. Note though, that extremely high
mass-loss rates are rare even in the Galaxy where they are thought to
represent the brief end of the most massive AGB stars. It is hardly
surprising that nothing similar has been found within our incomplete sample
of stars in small galaxies which are not generally believed to have an
intermediate mass population.

The measured mass-loss rates are all above the $\dot M_{\rm nuc}$ line. This has
important consequences for the evolution of these stars. Their evolution
will terminate when the mantle becomes depleted by mass-loss long before the
nuclear burning core can grow significantly. Similar conclusions were
reached about AGB stars in the LMC (van Loon et al. 1999b). This is independent on 
the value of $\psi$ we assume as the measured mass-loss rates are more than one 
order of magnitude higher than the nuclear burning rates.

The mass-loss rate versus bolometric luminosity diagram shows that most of
the stars fall within the range $4 \times 10^{-7} < \dot M < 6 \times
10^{-6}$ and
$-3.4 < M_{\rm bol} <-4.5$. Jackson et al. (2007) observed stars with the same
bolometric magnitudes and mass-loss rates in the low metallicity Local Group
galaxy WLM ([Fe/H]$\sim -1.13$). Stars with similar luminosities and
mass-loss rates have also been observed in the Large Magellanic Cloud (van
Loon et al. 1999b). These authors have pointed out, from dynamical
considerations following Gail \& Sedlmayr (1987), that dust-driven winds
below a few $10^{-6}$M$_{\odot}$yr$^{-1}$ with luminosities around a few
$10^3 $L$_{\odot}$ should not exist. Our observations thus disagree with those predictions.

\subsection{Comparison with mass-losing AGB stars in other galaxies}
\begin{figure}
\includegraphics[width=\columnwidth,clip=true]{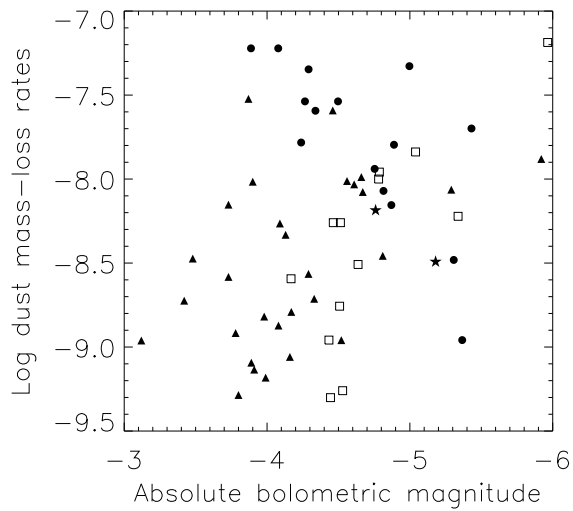}
\caption{\label{mass_loss_4gal} Dust mass-loss rates as a function of bolometric magnitude
 for stars in Sgr dSph (triangles), Fornax (stars), LMC (circles) and SMC (Squares).
}  
\end{figure}

To compare the mass-loss rates we measured in Fornax and Sgr dSph 
with mass-loss rates from AGB stars in the Magellanic Clouds (MCs) 
(thus in more metal-rich galaxies), we used the mass-loss rates 
derived by Groenewegen et al. (2007) for AGB stars in the MCs. 
To obtain an unbiased estimation of the bolometric magnitudes 
in these four galaxies, we determined it for the MCs sample in 
the same way we did for Fornax en Sgr dSph (using the JHK 
magnitudes and relation \ref{bol_cor}). The resulting relation 
between mass-loss rates and bolometric magnitude for these four galaxies is shown in Fig.\ref{mass_loss_4gal}.

 The first conclusion is that Fornax AGB stars, thus stars with very low metallicities,
 have quite small dust mass-loss rates for their luminosities. The smallest mass-loss rates are observed
 for  less luminous, more metal-rich AGB stars in Sgr dSph and the SMC, but it is likely 
that Fornax has lower mass-loss rate AGB stars too weak to be detected.

 For stars in the LMC the mass-loss rates are already high
 for a bolometric magnitude around $-$4 and then constant 
or even slightly decreasing. Note that two stars are outliers
 with M$_{\rm bol}$ around $-$5.5 and log(dust mass-loss rates)
 around $-$9 and $-$8.5. For stars in the SMC, the dust mass-loss
 rates increase steadily from  M$_{\rm bol}$$\sim$--4.5 to $\sim$--5.
 For M$_{\rm bol}$$\sim$--5, the observed mass-loss rates are of the
 same order of magnitude in the LMC and SMC. Stars in Sgr dSph show a
 similar behaviour, but the mass-loss rates begin to increase at
 M$_{\rm bol}$$\sim$--4.  The fact that the mass-loss rates are high
 for lower  luminosity in the LMC than in the other galaxies could
 be explained by the higher metallicity of the LMC with respect to the
 three other galaxies. It has been shown that at lower metallicity,
 stars become carbon-rich earlier on the AGB (e.g., Lagadec et al. 2007).
 So AGB stars become C-rich later in the LMC. The superwind might thus
 begin when the stars become carbon rich in the LMC, while the stars
 might be C-rich well before the onset of the superwind in the lower
 metallicity galaxies. The two Sgr dSph stars with the highest mass-loss
 rates and M$_{\rm bol}$$\sim$--4 (Sgr16 and Sgr18) would  thus be stars
 where the superwind phase necessary to explain the densities seen in typical
 planetary nebulae (Renzini 1981) has already begun.

If we compare the dust mass-loss rates for a given luminosity,
 then metallicity effects seem to appear. Indeed, for a given 
bolometric magnitude fainter than  --4.5, stars in the Sgr dSph and the SMC seem to have
 a smaller mass-loss rate than stars in the LMC. This 
could be due to the fact that Sgr dSph and the SMC have a smaller metallicity than
 the LMC and that at low luminosity less dust is formed in 
these galaxies than in the LMC, leading to a lower radiation pressure 
on dust grains and thus smaller mass-loss rates. Observations of a 
bigger sample of AGB stars in different galaxies are however needed to confirm this.
Indeed our MCs sample has been selected from the MSX catalogue (Egan et al. 2001) and is thus
 limited to stars brighter than 45 mJy in the MSX A band (8.3$\mu$m), preventing us to study the very early AGB.
 In Fornax we observed the brightest
 MIR stars while Sgr dSph stars were selected from NIR colors and sample the full AGB.  
Observations of fainter MCs stars
 in the MIR would enable us to study stars with fainter bolometric magnitude 
(with -4$<$M$_{\rm bol}$$<$-3) and compare it with some early AGB Sgr dSph stars we observed.

\section{Conclusions}

We have presented a study of mass-loss from AGB stars in the Sagittarius and
Fornax dwarf spheroidal galaxies using mid-infrared photometry obtained with
VISIR (VLT, ESO).

We have shown that the well known relation between $K-[12]$ colours and
mass-loss rates among Galactic stars is also valid for AGB stars in lower
metallicity galaxies. We used this relation and radiative transfer models to
estimate dust mass-loss rates for 15 stars in Sgr dSph and 2 stars in Fornax.

Our study showed that some AGB stars in the Sgr dSph  and Fornax galaxies are
losing mass. The estimated dust mass-loss rates are in the range
5$\times10^{-10}$ to $3\times10^{-8}$ M$_{\odot}$yr$^{-1}$ for the stars in
Sgr dSph and around 5$\times10^{-9}$ M$_{\odot}$yr$^{-1}$ for those in Fornax.
The values obtained with the two different methods are of the same order of
magnitude. These mass-loss rates are higher than the nuclear burning rates,
so these stars will terminate their AGB evolution by the depletion of their
stellar mantles, before their cores can grow significantly.

Using previous work to get an estimation of the dust-to-gas mass ratio for
the stars observed we found that most of them had very low mass-loss rates
for their luminosity. This is in contradiction with theoretical predictions (see e.g Gail \& Sedlmayr 1987).


\section*{Acknowledgments}
E.L. acknowledges support from a PPARC rolling grant.
We thank the referee, Joris Blommaert, for his useful comments that helped improving the quality of the paper.

\label{lastpage}
\end{document}